\documentstyle[sprocl,psfig]{article}

\def\be{\begin{equation}}
\def\ee{\end{equation}}
\def\bea{\begin{eqnarray}}
\def\eea{\end{eqnarray}}
\newcommand{\ab}{{_{{\scriptscriptstyle A,B}}}}
\newcommand{\da}{{{\delta}_{{\scriptscriptstyle A}}}}
\newcommand{\done}{{{\delta}_{{\scriptscriptstyle 1}}}}
\newcommand{\dtwo}{{{\delta}_{{\scriptscriptstyle 2}}}}
\newcommand{\ma}{{m_{{\scriptscriptstyle A}}}}
\newcommand{\mone}{{m_{{\scriptscriptstyle 1}}}}
\newcommand{\mtwo}{{m_{{\scriptscriptstyle 2}}}}
\newcommand{\pab}{{p_{{\scriptscriptstyle AB}}}}
\newcommand{\phionetwo}{{\Phi_{{\scriptscriptstyle 12}}}}
\newcommand{\ponetwo}{{p_{{\scriptscriptstyle 12}}}}
\newcommand{\rone}{{{\vec r}_{{\scriptscriptstyle 1}}}}
\newcommand{\ro}{\scriptscriptstyle {\rho}}
\newcommand{\rtwo}{{{\vec r}_{{\scriptscriptstyle 2}}}}

\newcommand{\rv}{{\vec r}}
\newcommand{\ra}{{{\rv}_{{\scriptscriptstyle A}}}}
\newcommand{\rb}{{{\rv}_{{\scriptscriptstyle B}}}}
\newcommand{\rh}{{\hat r}}
\newcommand{\rhone}{{{\hat r}_{{\scriptscriptstyle 1}}}}
\newcommand{\rhtwo}{{{\hat r}_{{\scriptscriptstyle 2}}}}

\newcommand{\rha}{{{\hat r}_{{\scriptscriptstyle A}}}}
\newcommand{\rhb}{{{\hat r}_{{\scriptscriptstyle B}}}}
\newcommand{\RoA}{{{\rho}_{{\scriptscriptstyle A}}}}
\newcommand{\Rone}{{{\rho}_{{\scriptscriptstyle 1}}}}
\newcommand{\Rotwo}{{{\rho}_{{\scriptscriptstyle 2}}}}
\newcommand{\sA}{{s_{{\scriptscriptstyle A}}}}
\newcommand{\sB}{{s_{{\scriptscriptstyle B}}}}
\newcommand{\sone}{{s_{{\scriptscriptstyle 1}}}}
\newcommand{\stwo}{{s_{{\scriptscriptstyle 2}}}}

\newcommand{\se}{{\sigma_{{\scriptscriptstyle 8}}}}
\newcommand{\va}{{{\vec v}_{{\scriptscriptstyle A}}}}
\newcommand{\vs}{{v_{{\scriptscriptstyle 12}}}}

\newcommand{\vv}{{{\vec v}_{{\scriptscriptstyle 12}}}}
\newcommand{\vt}{{{\tilde v}_{{\scriptscriptstyle 12}}}}
\newcommand{\vone}{{{\vec v}_{{\scriptscriptstyle 1}}}}
\newcommand{\vtwo}{{{\vec v}_{{\scriptscriptstyle 2}}}}
\newcommand{\xb}{\bar{\xi}}

\newcommand{\xbb}{\bar{\hspace{-0.08cm}\bar{\xi}}}

\begin{document}

\title{Measuring $\Omega$ with galaxy streaming velocities}

\author{P. G. Ferreira$^{1,2}$,
M. Davis$^{3}$,
H. A. Feldman$^{4,5}$,
A. H. Jaffe$^{3}$,
R. Juszkiewicz$^{4,5,6,7}$,
 }

\address{
$^{1}$ Theory Group, CERN, CH-1211, Geneve 23, Switzerland\\
$^{2}$ CENTRA, Instituto Superior Tecnico, Lisboa 1096 Codex, Portugal\\
University of California, Berkeley, CA94720, USA\\
$^{4}$D{\'e}partement de Physique Th{\'e}orique, Universit{\'e}
de Gen{\`e}ve, CH-1211 Gen{\`e}ve, Switzerland\\
$^{5}$Department of Physics and Astronomy, University of Kansas,
Lawrence, KS 66045\\
$^{6}$
On leave from
Copernicus Astronomical Center, 00-716 Warsaw, Poland\\
$^{7}$Joseph Henry Laboratories,
Princeton University,
Princeton NJ 08544}


\maketitle\abstracts{
The mean pairwise velocity of galaxies,  
$\vs$ has traditionally been estimated from the
redshift space galaxy correlation function. This method is notorious
for being highly sensitive
to the assumed model of the pairwise velocity
dispersion. Here we propose an
alternative method to estimate $\vs$ directly from peculiar velocity
samples, which contain redshift-independent distances as well as galaxy
redshifts.
This method
can provide an estimate of $\Omega^{0.6}\se^2$
for a range of $\se$
where $\Omega$ is the cosmological density parameter, while $\se$
is the standard normalization for the power spectrum of density
fluctuations. We demonstrate how to measure this quantity from
realistic catalogues and identify the main sources of bias and errors
}
  
\section{A model of $v_{12}(r)$}

In this presentation we report on the the possibility of using the
``mean tendency of well-separated galaxies to approach each other''
\cite{peebles80} to measure the cosmological
density parameter, $\Omega$. The statistic we consider\cite{ferreira98} is the mean relative
pairwise velocity of galaxies, $\vs$. It was introduced
in the context of the BBGKY theory \cite{davis77},
describing the dynamical evolution of a collection of particles
interacting through gravity. In this discrete picture,
$\vv$ is defined as the mean value of the peculiar
velocity difference of a particle pair at separation ${\vec r}$. 
In the fluid limit, its analogue is the
pair-density weighted relative velocity
\cite{fisher94,rj98a},
\begin{eqnarray}
\vv(r) \; = \; \langle \, \vone - \vtwo \, \rangle_{\ro} \; = \;
{{ \langle(\vone -
\vtwo ) (1 + \done)(1 +\dtwo ) \rangle}
\over {1 \; + \; \xi(r)}} \; ,
\label{def}
\end{eqnarray}
where $\, \va \,$ and $\, \da = \RoA / \langle \rho \rangle - 1 \,$
are the peculiar velocity and fractional density contrast
of matter at a point $\ra$, $r=|\rone - \rtwo|$, and $\xi(r) =
\langle\done\dtwo\rangle$ is the two-point correlation function.
The pair-weighted average, $\langle\cdots\rangle_{\ro}$,
differs from simple spatial averaging, $\langle\cdots\rangle$,
by the weighting factor
$\Rone\Rotwo \,\langle \Rone\Rotwo\rangle^{-1}$,
proportional to the number-density of particle pairs.
In a recent letter \cite{rj98}, one of us has shown that an excellent approximation to $\vs$ is given by 
\begin{eqnarray}
\vs (r) \; &=& \; - \, {\textstyle{2\over 3}} \, H r f \,
\xbb (r)[1 + \alpha \; \xbb (r)] \;,
\label{2nd-order}\\
\xb(r) \; &=& \; (3/r^3)\, \int_0^r \, \xi(x) \, x^2 \, dx \; 
\equiv \; \xbb(r) \, [\, 1 + \xi(r) \, ]
\label{xb}
\end{eqnarray}
Here
$\alpha$ is a parameter, which depends on the logarithmic slope
of $\xi(r)$, while
$f = d \, \ln D/d \ln a$, with
$D(a)$ being the standard linear growing mode
solution and $a$ -- the cosmological
expansion factor. 
Finally, $H = 100\;h\;\mbox{km}\;\mbox{s}^{-1}\;\mbox{Mpc}$ is 
the present value of the Hubble constant.
This approximate solution of the pair conservation equation,
 accurately
reproduces results of high resolution N-body simulations
in the entire dynamical range \cite{rj98}.

If we restrict ourselves to $\, r = 10 h^{-1}\, \mbox{Mpc} \,$,
one can use the APM catalogue of galaxies \cite{efst96} for an
estimate of galaxy correlation function, $\xi=(r/r-0)^{-\gamma}$ ; 
the slope at the separation considered
is $\gamma=1.75\pm0.1$ (the errors we quote are conservative).
One obtains then 
\begin{equation}
\vs(10 h^{-1}\;\mbox{Mpc}) \; =
\; - \, 605 \, {\se}^2 \Omega^{0.6}
\,(1 + 0.43\se^2)\,/(1+ 0.38{\se}^2)^2 \, {\rm km/s}
\; .
\label{10mpc}
\end{equation}
The above relation shows that at $r = 10 h^{-1}\;\mbox{Mpc}$, $\vs$ is
almost entirely determined by the values of two parameters: $\se$ and
$\Omega$. It is only weakly dependent on $\gamma$. 
The uncertainties in the
observed $\gamma$ lead to an error in Eq.~\ref{10mpc} of less than
$10\%$ for $\se\le 1$.

\section{The estimator}

Since we observe only the line-of-sight component of the
peculiar velocity, $\sA = \ra\cdot \va/r \equiv 
\rha\cdot \va$, rather than the
full three-dimensional velocity $\va$, it is not possible
to compute $\vs$ directly. Instead, we propose to use the
mean difference between radial velocities of a pair of
galaxies,
$\langle \, \sone - \stwo \, \rangle_{\ro} \; = \;
\vs \, \rh\cdot(\rhone + \rhtwo)/2$,
where $\rv = \rone - \rtwo$.
To estimate $\vs$, we use the simplest least squares techniques,
which minimizes the quantity
$\chi^2(r) \; = \; \sum_{\ab} \, \left[ (\sA - \sB) - \pab
\,\vt(r)/2 \, \right]^2 \; \;$,
where $\pab \equiv \rh \cdot (\rha + \rhb)$ and
the sum is over all pairs at fixed separation $r = |\ra - \rb|$.
The condition $\partial \chi^2 / \,\partial\vt = 0$ implies
\begin{eqnarray}
\vt (r) \; = \; {
{2 \sum \, (\sA - \sB)\, \pab }\over
{\sum \pab^2}} \;\; .
\label{estimator}
\end{eqnarray}
The above expression is a sum over positive quantities
and so is stable. This estimator is appropriate to be
applied to a point process which will sample an underlying continuous
distribution. The sampling is quantified in terms of the selection
function, $\phi(\rv)$. The continuum limit of Eq.~\ref{estimator}
is then
\begin{eqnarray}
\vt (r)=\frac{2 \int d \mone \,
d \mtwo \, \phionetwo \,
(\sone - \stwo)\ponetwo}
{\int d \mone \,
d \mtwo \, \phionetwo \,
\ponetwo^2} \; ,
\label{estcont}
\end{eqnarray}
with $d \ma = \RoA~d^3 \ra$, and
a two-point selection function given by
$\phionetwo \; = \; \delta_D(|{\rone - \rtwo}| - r)
\, \phi(\rone) \, \phi(\rtwo) \;$,
where $\delta_D$ is the Dirac delta function.
For ease of notation we shall denote the denominator
in Eq.~\ref{estcont} by $W(r)$.  If
we take the ensemble average of Eq.~\ref{estcont} we
 find that $\langle \vt(r) \rangle=\vs (r)$.  Note that,
unlike the estimators for the velocity correlation tensor proposed in
\cite{gorski89}, the ensemble average of the estimator is $\vs (r)$
independent of
the selection function. For an isotropic selection function this
estimator is insensitive to systematic effects such as a bulk flow,
large scale shear and small scale random velocities
(as one might expect from virialized objects).

\section{Biases and errors}

How unbiased is this estimator? We have applied our
statistic to mock catalogues extracted from N-body simulations of a
dust-filled universe with $\Omega=1$ and $P(k) \propto 1/k$.

\begin{figure}[t]
\hspace{0.7in}
\psfig{figure=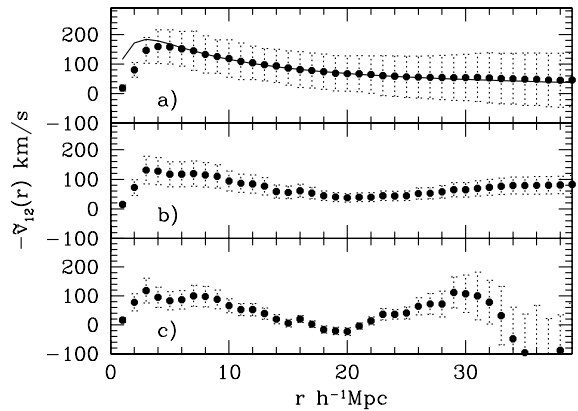,width=3.3in}
\vskip -0.7in
\caption{$\vt(r)$ (points) and its variance (dashed lines)
evaluated from mock catalogues: A) random
observers with a deep  selection function compared to $\vs (r)$
evaluated from Eq.~\ref{2nd-order} (solid line); b)
A fixed observer with a deep selection function; c) a fixed observer
with a shallow selection function. The variance is estimated
from the scatter over 20 (a) or 9 (b,c) mock catalogues.
\label{fig2}}
\end{figure}
In Figure~\ref{fig2}(a) we plot $\vt(r)$ with one standard
deviation calculated with 20 mock catalogues extracted with a
deep selection function. Each catalogue has a
different observation position within the simulation volume and so an
average over this set should resemble a true ensemble average. The mean
is consistent with what one would expect from a
direct calculation with Eq.~\ref{2nd-order} (which is plotted in
Figure~\ref{fig2}(a) as a solid line).  We have also performed this
analysis without collapsing the cores; the results changed by very
little. 

We repeat this
calculation for a set of 9 catalogues all constructed from the same
observation point  using a deep (Figure~\ref{fig2}b) or shallow 
(Figure~\ref{fig2}c) selection
function to randomly sample a fraction of galaxies within the
simulation box. 
The variance
in $\vt(r)$ is now solely due to finite sampling (``shot noise'');
for catalogues with 2000 to 3000 galaxies we expect the
variance to be $\sqrt{2}$--$\sqrt{3}$ times larger.
We find that a shallow selection function
changes the functional form, or slope, of the mean,
making it a more sharply decreasing
function of $r$ than the ensemble average.  It is therefore crucial when
analyzing a catalogue to restrict oneself to scales much smaller than
the effective cutoff scale of the selection function.

Errors in distance measurements will naturally affect our
results; the best estimators
use empirical correlations between intrinsic properties of
the galaxies and luminosities and 
lead to log-normal errors in the estimated distance of around
$20\%$. These errors will naturally lead to biases in cosmological
estimators involving distance measurements and peculiar velocities and
are generically called Malmquist bias. 
We shall model our errors assuming a Tully--Fisher law which
resembles that inferred from the Mark III catalogue.  To correct for
Malmquist bias we use the prescription put forward 
in \cite{landy92}.

\begin{figure}[t]
\hspace{0.7in}{
\psfig{figure=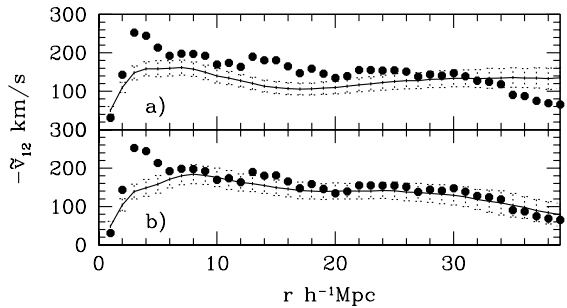,width=3.3in}}
\vskip -0.7in
\caption{$\vs (r)$  and its variance evaluated from 100 mock catalogues
with errors (described in the text) and the full selection function. The
solid points are the  $\vt$ of the error-free simulation seen from the
same observation point,
the solid line is the mean and dashed lines are the 1 $\sigma$. a)
uncorrected distances; b) distances corrected for Malmquist bias
\label{fig3}}
\end{figure}

In Figure~\ref{fig3}(a), we plot the results for the uncorrected simulations;
Malmquist errors systematically lower the values of $\vt$
on small scales while enhancing its amplitude on large scales
(where the effect should be more dominant).
However in Figure~\ref{fig3}(b) we show
that with the correction for general Malmquist errors to the distance
estimator, it is possible to overcome this discrepancy. The 1-$\sigma$
errors now encompass the true $\vt$ over a wide range of
scales.

\section{Conclusions}

In this contribution we report on a recent proposal to estimate the
mean pairwise streaming velocities of galaxies
directly from peculiar velocity samples. 
We identified three possible sources of systematic errors in
estimates of $\vs$ made directly from radial peculiar velocities
of galaxies. We also found ways of
reducing these errors; these techniques were successfully tested
with mock catalogues. The potential sources of errors and their
proposed solutions can be summarized as follows.

(1) On the theoretical front, assuming a linear theory
model of $\vs(r)$ at $r \approx
10h^{-1}\;\mbox{Mpc}$ can introduce a considerable systematic error in
the resulting estimate of $\se^2\Omega^{0.6}$.
For example, if $\se = 1$ using the linear prediction for
$\vs$ at $r = 10 h^{-1}\;\mbox{Mpc}$ would introduce a 25\%
systematic error (see eq.~[\ref{10mpc}]).
We solve this problem by using  a nonlinear expression
for $\vs$.\cite{rj98}

(2) On the observational front,
a shallow selection function induces a
large covariance between $\vt$ on different scales. This
must be taken into consideration by measuring $\vt(r)$ only
on sufficiently small scales. A rule of thumb is that for
estimating $\vt$ at $10h^{-1}\;\mbox{Mpc}$, the selection function
should be reasonably homogeneous out to at least $30h^{-1}\;\mbox{Mpc}$.

(3) Finally, care must be taken with generalized Malmquist bias due to
log-normal distance errors; these induce a systematic error in $\vt$.
We have shown that, under certain assumptions about selection and
measurement errors, the method of Landy $\&$ Szalay\cite{landy92} for corrected
distance estimates allows one to recover the true $\vt$.  Naturally,
this particular correction must be addressed on a case-by-case basis,
given that different data sets will have different selection criteria
and correlations between galaxy position and measurement errors.

In a future publication we shall analyze the Mark III \cite{willick97}
 and the SFI \cite{dacosta96} catalogues of
galaxies with this in mind.

\section*{Acknowledgments}
We thank the organizers for an enjoyable meeting. 
We thank Jonathan Baker, Stephane
Courteau, Luis da Costa and Jim Peebles,
for useful comments and suggestions.  
This work was supported by
NSF, NASA, 
the Poland-US M. Sk{\l}odowska-Curie Fund,
the Tomalla Foundation and JNICT.

\end{document}